\documentclass[final]{aipproc}
\layoutstyle{6s}

\newcommand{\al}{\alpha}
\newcommand{\bb}{\beta}

\newcommand{\ben}{\begin{eqnarray}}
\newcommand{\een}{\end{eqnarray}}
\newcommand{\be}{\begin{equation}}
\newcommand{\ee}{\end{equation}}
\newcommand{\n}{\label}
\newcommand{\no}{\noindent}

\newcommand{\ga}{\gamma}
\newcommand{\ro}{\rho}

\newcommand{\bn}{\begin{equation}\label}

\begin{document}
\title{Holographic dark energy interacting with dark matter}

\author{M\'onica I. Forte}{address={Departamento de F\'isica, Facultad de ciencias Exactas y Naturales, Universidad de Buenos Aires, 1428 Buenos Aires, Argentina},email={monicaforte@fibertel.com.ar}}
\author{Mart\'{i}n G. Richarte}{address={Departamento de F\'isica, Facultad de ciencias Exactas y Naturales, Universidad de Buenos Aires, 1428 Buenos Aires, Argentina},email={martin@df.uba.ar}}

\begin{abstract}We investigate a spatially flat Friedmann-Robertson-Walker (FRW) cosmological model with cold dark matter coupled to a dark energy which is given by the modified holographic Ricci cutoff. The interaction used is linear in both dark energy densities, the total energy density and its derivative. Using the statistical method of $\chi^2$-function for the Hubble data, we obtain  $H_0=73.6km/sMpc$,  $\omega_s=\gamma_s -1=-0.842$ for the asymptotic equation of state and   $ z_{acc}= 0.89 $. The estimated values of  $\Omega_{c0}$   which fulfill the current observational bounds  corresponds to a dark energy density varying in the range  $0.25R < \ro_x < 0.27R$.
\end{abstract}

\maketitle

\bibliographystyle{plain}

\section{Introduction}


Many different observational sources such as the Supernovae Ia \cite{astro-ph/9805201}-\cite{astro-ph/9812133}, the large scale structure from  the Sloan Digital Sky survey \cite{arXiv:0707.3413} and the cosmic microwave background anisotropies \cite{arXiv:1001.4538}  have corroborated that our universe is currently undergoing an accelerated phase. The cause of this behavior has been attributed to a mysterious component called dark energy and several candidates have been proposed to fulfill this role. For example, a positive cosmological constant $\Lambda$, explains very well the accelerated behavior but it has a deep  mismatch with the  theoretical value predicted by the quantum field theory. Another issue of debate refers to the coincidence problem, namely: why the dark energy and dark matter energy densities happen to be of the same order precisely today. In order to overcome both problems, it has proposed a dynamical framework in which the dark energy varies with the cosmic time. This proposal  has led to a great variety of dark energy models such as quintessence \cite{astro-ph/9807002}, exotic quintessence \cite{arXiv:0706.4142},  N--quintom \cite{arXiv:0811.3643}  and the holographic dark energy (HDE) models \cite{hep-th/0403127} based  in an application of the holographic principle to the cosmology.    According to this principle, the entropy of a system does not scale with its volume but with its surface area and so in cosmological context will set an upper bound on the entropy of the universe \cite{hep-th/9806039}. In \cite{hep-th/9803132}  it has been suggested that in quantum field theory a short distance cut-off is related to a long distance cut-off (infra-red cut-off L) due to the limit set by the formation of a black hole. Further, if  the quantum zero-point energy density caused by a short distance cut-off is taken as the dark energy density in a region of size  L, it should not exceed black hole mass of the same size, so $\rho_{\Lambda}=3c^{2}M^{2}_{~P}L^{-2}$, where $c$ is a numerical factor. In the cosmological context, the size L is usually taken  as the large scale of the universe, thus Hubble horizon, particle horizon, event horizon or generalized IR cutoff. Between all the interesting holographic dark energy models proposed so far,  here  we focus our attention on a modified version of the well known Ricci scalar cutoff proposed in \cite{arXiv:0810.3663}.  Besides, there could be a hidden non-gravitational coupling between the dark matter and dark energy without violating current observational constraints and thus it is interesting to develop ways of testing an interaction in the dark sector. Interaction within the dark sector has been studied mainly as a mechanism to solve the coincidence problem.   We will consider an exchange of energy or interaction between dark matter and dark energy  which  is a linear combination of the dark energy density $\ro_x$,  total energy density $\ro$, dark matter energy density $\ro_c$, and the first derivate of the total  energy density  $\ro'$ as has been studied in \cite{arXiv:0911.5687}.


\section{The interacting model}


In a FRW background, the Einstein equation for a model of cold dark matter and modified holographic Ricci dark energy, having energy densities $\rho_c$ and $ \rho_x $ respectively, reads 

\be
\n{Friedmann}
3H^2=\rho, \qquad  \rho_x=\left(2\dot H + 3\alpha H^2\right)/\Delta,\ee

\no where $\alpha$, $\beta$ are constants and $\Delta=\alpha -\beta$. In terms of the variable  $\eta = 3\ln(a/a_0)$, the  compatibility between the global conservation equation 

\be
\n{cons}
\rho ' = d\rho/d\eta= -\rho_c-(1+w_x)\rho_x, 
\ee

\no and the modified holographic Ricci dark energy 

\be
\n{comp}
\rho ' = -\alpha\rho_c-\beta\rho_x,
\ee

\no namely, $(\ga_c\ro_c + \ga_x\ro_x)=(\al\ro_c + \bb\ro_x)$, gives a relation between the equation of state of the dark components $w_x$ and the ratio $r= \rho_c/\rho_x$

\ben
\n{relacEos}
w_x=(\alpha  - 1)r+\beta-1.
\een

Solving the Eq. (\ref{comp}) along with $\ro=\ro_c+\ro_x$, we get the dark energy densities in terms of $\ro$ and $\ro'$
 
\ben
\n{dens}
\rho_c=-(\beta\rho+\rho')/\Delta,  \qquad   \rho_x=(\alpha\rho+\rho')/\Delta.
\een

We introduce the interaction between both dark components through the term $Q$ by splitting the Eq.(\ref{cons}) into  $\rho'_c+\rho_c = - Q$  and $\rho'_x+(1+w_x) \rho_x =  Q$. Also, we present a modified interaction $Q_M$ by splitting the Eq. (\ref{comp}) into $\rho'_c+ \alpha \rho_c = - Q_M $ and  $ \rho'_x+ \beta \rho_x  =   Q_M $, meaning that both pictures are related by means of the expression $Q = Q_M - (1-\alpha) \rho_c = Q_M+(1+w_x-\beta)\rho_x$. Then, differentiating $\rho_c$ or $\rho_x$ in (\ref{dens}) and using the expression of  $ Q_M $  allow us to obtain a second order differential equation for the total energy density $\rho$  \cite{arXiv:0911.5687}

\be
\n{2doOrden}
\rho''+(\alpha+\beta)\rho'+\alpha\beta\rho = Q_M \Delta.
\ee

Solving Eq. (\ref{2doOrden}) for a given interaction $Q_{M}$, we obtain the total energy density $\ro$ and the dark matter and dark energy densities $\ro_{c}$, $\ro_{x}$ after using Eq. (\ref{dens}).

The general interaction $ Q_M $ linear in $\ro_{c}$, $\ro_{x}$, $\ro$, and $\ro'$ has been introduced in \cite{arXiv:0911.5687},

\ben
Q_M= c_1 \frac{(\gamma_s - \alpha)(\gamma_s-\beta)}{\Delta\gamma}\,\rho + c_2 (\gamma_s-\alpha)\rho_c - c_3 (\ga_s -\bb)\ro_x -c_4 \frac{(\ga_s - \al)(\ga_s-\bb)}{\ga_s\Delta\ga}\,  \rho',
\label{ql}
\een

\no where $\ga_s$ is constant and  the coefficients $c_{i}$ fulfill the condition $c_{1}+c_{2}+c_{3}+c_{4}=1$.  

Now, using Eqs. (\ref{dens}) we rewrite the interaction (\ref{ql}) as a linear combination of $\ro$ and $\ro'$, 

\ben
Q_M=\frac{u\ro+\ga^{-1}_{s}[u-(\ga_s-\al)(\ga_s-\beta)]\ro'}{\Delta\ga}, 
\label{qfinal}
\een

\no where $u=c_1(\ga_s -\al)(\ga_s-\beta)-c_{2}\beta(\ga_s-\al)-c_{3}\al(\ga_s-\beta)$.  Replacing the interaction (\ref{qfinal}) into the source equation (\ref{2doOrden}), we obtain 

\begin{equation}
\ro'' + (\ga_s+ \ga^{+})\ro'+ \ga_{s}\ga^{+}\ro=0.
\label{se}
\end{equation}

\no where the  roots of the characteristic polynomial associated with the second order linear differential equation (\ref{se}) are  $\ga_{s}$ and $\ga^{+}=(\beta\al -u)/{\ga_s}$. In what follows, we adopt $\ga^{+}=1$ for mimicking the dust-like behavior of the universe at early times. In that case, the general solution of (\ref{se}) is   $\rho=b_1a^{-3\gamma_s}+b_2a^{-3}$  from which we obtain

\ben
\n{rhocc}
\rho_c=\frac{(\gamma_s-\beta)b_1a^{-3\gamma_s}+(1-\beta)b_2a^{-3}}{\Delta},\\
\n{rhocx}
\rho_x=\frac{(\alpha-\gamma_s)b_1a^{-3\gamma_s}+(\alpha-1)b_2a^{-3}}{\Delta}.
\een

Interestingly, Eqs. (\ref{rhocc}) and (\ref{rhocx})  tell us that  the interaction (\ref{qfinal}) seems to be a good candidate for alleviating the cosmic coincidence problem because the ratio $\Omega_{c}/\Omega_{x}$ becomes bounded for all times.

\section{Observational constraints}

In terms of the transition redshift $z_{acc}$, the actual Hubble factor $H_0$ and the $\gamma_s$ parameter, we obtain the Hubble function

\be
\n{H}
H(z)= \frac{H_0(1+ z_{acc})^{3/2}} {\sqrt{2-3\gamma_s+(1+ z_{acc})^{3(1-\gamma_s)}}}\sqrt{\frac{(1+z)^{3\gamma_s}}{(1+z_{acc})^{3\gamma_s}}+(2-3\gamma_s)\frac{(1+z)^3}{(1+z_{acc})^3}}
\ee

\begin{figure}
  \resizebox{20pc}{!}{\includegraphics{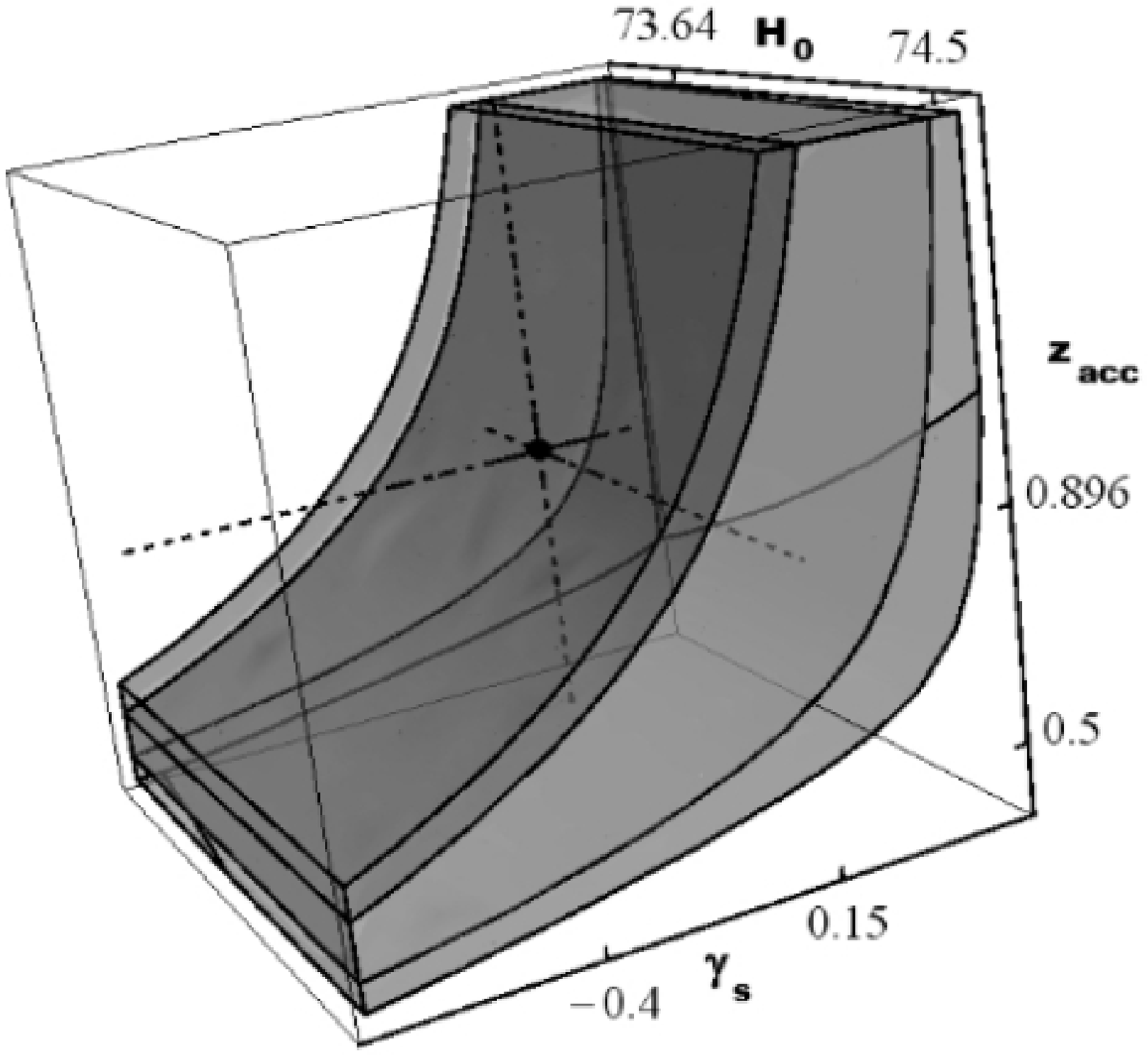}}
  \resizebox{18pc}{!}{\includegraphics{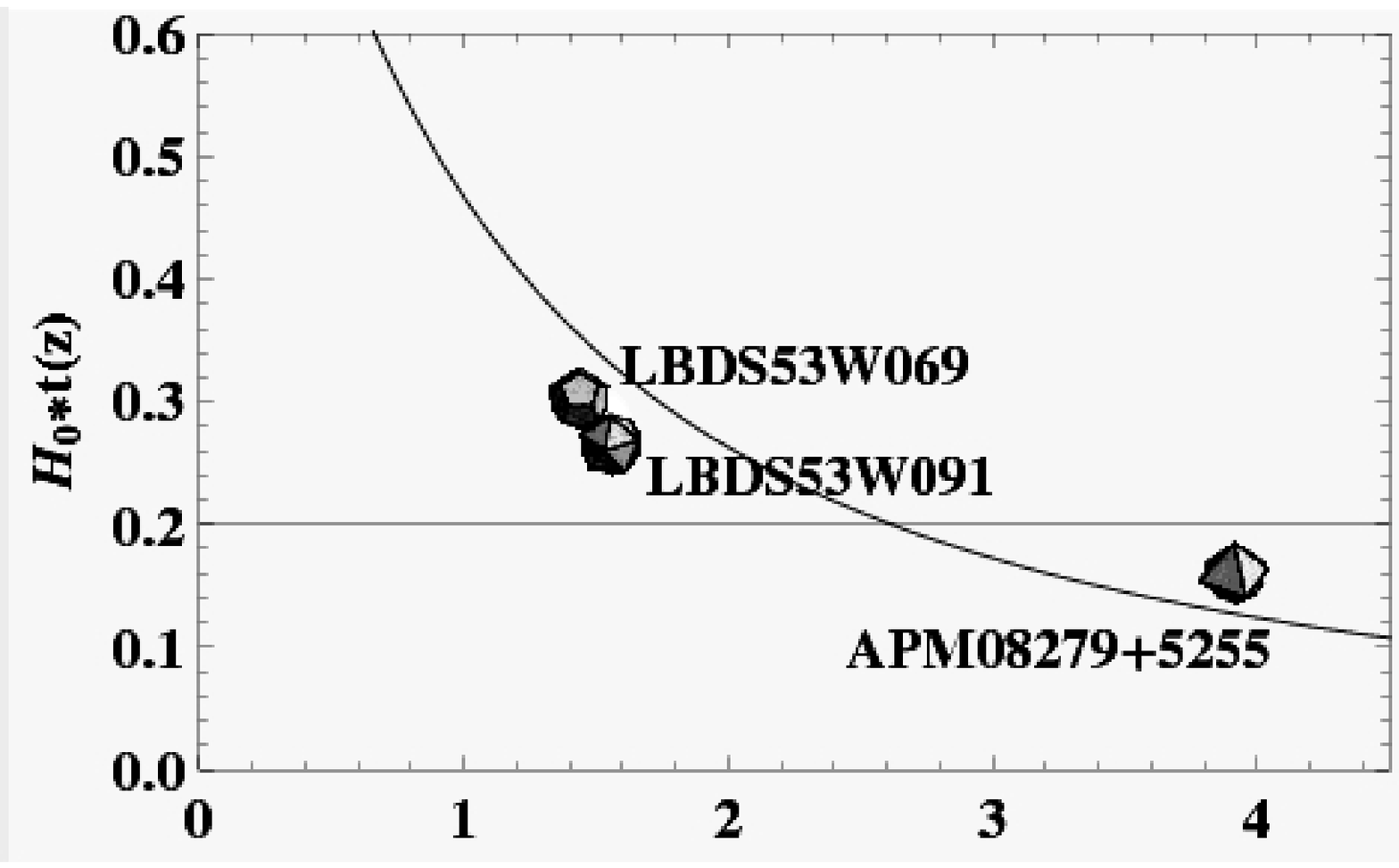}}
\caption{Left panel: Three-dimensional regions of confidence $1\sigma$ and $2\sigma$ for  $H_0$, $\gamma_s=1+w_s$  \  and  $z_{acc}$ parameters. Right panel: Parametric curve of cosmological time $t(z)$ drawn in units of $H_0^{-1}$ for the best values $z_{acc}=0.89$, $H_0=73.6km/sMpc$  \ and  $\gamma_s=0.158$ with (\ref{tiempo}).}

\label{figura}\end{figure}

We apply the $\chi^{2}$--statistical method to the Hubble data listed in \cite{Stern:2009ep} for  constraining the cosmological parameters of the Hubble function (\ref{H}). The three-dimensional confidence regions $1\sigma$ and $2\sigma$ are shown in the left panel of Fig.\ref{figura} where the sphere indicates the best fit values $z_{acc}=0.89$, $H_0=73.6km/sMpc$  \ and  $\gamma_s=0.158$.

A feasible model of the dark sector has dark components with positive definite energy densities, accelerated expansion and non
phantom dark energy. These requirements are fulfilled when    $b_1$ and $b_2$  are positive constants, which correspond to  $\alpha \geq 1$ and $ 0 \leq \beta < 2/3$.  It is possible to infer that in the case of $\alpha = 4/3$ and $ \beta < 0.1$, that is $0.25R < \rho_x < 0.27R$, the model behaves better than the one based on the Ricci scalar.

\vskip0.2cm 
\subsection{The crisis of the age}

The age of the universe in units of $H_0^{-1}$ can be obtained as a function of the redshift $z$ with the expression 
\be
\n{tiempo}
t(z)= \int_z^{\infty} \left((1+\stackrel{-}{z})H(\stackrel{-}{z})\right)^{-1}d\stackrel{-}{z}
\ee
We can see in the right panel of Fig.\ref{figura} that our modified holographic model with linear interaction works very well at least for values of redshift  lower than $z \sim 3.5$. The use of a nonlinear interaction likely includes more ancient objects such as $APM08279+5255$.
 
\section{Conclusions}
 We have examined a modified holographic Ricci dark energy  coupled  with cold dark matter and found that this scenario describes satisfactorily the evolution of both dark components. 
We have shown that the compatibility between the modified and the global conservation equations constraints the equation of state of the dark  energy component. From the observational point of view we have  obtained the best fit values of the cosmological parameters $z_{acc}=0.89$, $H_0=73.6km/sMpc$  \ and  $\gamma_s=0.158$   with   a $\chi^{2}_{dof}=0.761 < 1$ per degree of freedom. The $H_{0}$ value is in agreement with the reported in the literature \cite{Riess:2009pu} and the critical redshift $z_{acc}=0.89$ is consistent with BAO and CMB data \cite{Li:2010da}. We have found that the age crisis at high redshift cannot be alleviated so it will need another kind of interaction.

\begin{theacknowledgments}

M.I. Forte is grateful for the invitation and for financial support provided by I GAC. M. G. Richarte is partially supported by CONICET.
The authors are grateful with the Prof.  L.P. Chimento for a careful reading of the manuscript. 

\end{theacknowledgments}

\end{document}